\documentclass[aps,prb,twocolumn,superscriptaddress,showpacs]{revtex4}
\usepackage{graphicx}

\begin{document}

\title{Transport and localization in periodic and disordered graphene
superlattices.}
\author{Yury P. Bliokh}
\affiliation{Advanced Science Institute, The Institute of Physical and
Chemical Research
(RIKEN), Wako-shi, Saitama 351-0198, Japan}
\affiliation{Physics Department, Technion-Israel Institute of Technology,
Haifa 32000,
Israel}
\author{Valentin Freilikher}
\affiliation{Advanced Science Institute, The Institute of Physical and
Chemical Research
(RIKEN), Wako-shi, Saitama 351-0198, Japan}
\affiliation{Jack and Pearl Resnick Institute of Advanced Technology,
Department of
Physics, Bar-Ilan University, Ramat-Gan 52900, Israel}
\author{Sergey Savel'ev}
\affiliation{Advanced Science Institute, The Institute of Physical and
Chemical Research
(RIKEN), Wako-shi, Saitama 351-0198, Japan}
\affiliation{Department of Physics, Loughborough University, Loughborough
LE11 3TU,
United Kingdom}
\author{Franco Nori}
\affiliation{Advanced Science Institute, The Institute of Physical and
Chemical Research
(RIKEN), Wako-shi, Saitama 351-0198, Japan}
\affiliation{Center for Theoretical Physics, Department of Physics, CSCS,
University of
Michigan, Ann Arbor, Michigan 48109-1040, USA}

\begin{abstract}
We study charge transport in one-dimensional graphene superlattices created
by applying layered periodic and disordered potentials. It is shown that the
transport and spectral properties of such structures are strongly
anisotropic. In the direction perpendicular to the layers, the eigenstates
in a disordered sample are delocalized for all energies and provide a
minimum non-zero conductivity, which cannot be destroyed by disorder, no
matter how strong this is. However, along with extended states, there exist
discrete sets of angles and energies with exponentially localized
eigenfunctions (disorder-induced resonances). Owing to these features, such
samples could be used as building blocks in tunable electronic circuits. It
is shown that, depending on the type of the unperturbed system, the disorder
could either suppress or enhance the transmission. Remarkable
properties of the transmission have been found in graphene systems built of
alternating p-n and n-p junctions. The mean transmission coefficient has
anomalously narrow angular spectrum, practically independent of the
amplitude of the fluctuations of the potential. To better understand the
physical implications of the results presented here, most of these have been
compared with the results for analogous electromagnetic wave systems. Along with
similarities, a number of quite surprising differences have been found.
\end{abstract}

\maketitle

\section{Introduction.}

\bigskip

The exploration of graphene is nowadays one of the most animated areas of
research in condensed matter physics (see, e.g., Refs.\cite%
{Katsnelson2,Geim,periodic-1}). Its unique properties not only arouse pure
scientific curiosity but also suggest possible practical applications{\LARGE
.} More in-depth studies of graphene continuously bring about more
counterintuitive discoveries. Examples are plentiful. Suffice to mention a
novel integer quantum Hall effect \cite{dis-exp-1, Castro}, total
transparency of any potential barrier for normally-incident electrons/holes
\cite{Katsnelson1} (in analogy with the Klein paradox \cite{Klein}), and the
recently predicted focusing of electron flows by a rectangular potential
barrier \cite{Cheianov} (an analog of the Veselago lens \cite{Veselago,
Pendry}). Of even greater surprise are the properties of disordered graphene
systems \cite{disorder-1, disorder-2,
Titov-1,San-Jose,disorder-3,disorder-4,dis-exp-2,dis-exp-3}. The latest
results, both theoretical and experimental{\LARGE ,} led to the amazing
conclusion that there is no localization in disordered graphene, even in the
one-dimensional situation, i.e., when the random potential depends only on
one coordinate. In this paper, we show that this conclusion, if taken
unreserved, could be misleading. We demonstrate here that a well-pronounced
localization can take place in graphene, i.e., there could exist a
(quasi)-discrete spectrum with exponentially localized eigenfunctions \cite%
{remark-1}. This localization can occur even though disorder can never make
a graphene sample a complete insulator, and there is always a minimal
residual conductivity (an indication of delocalization). In this paper, the
charge transport in periodically and randomly layered graphene structures is
studied and analogies with the propagation of light in layered dielectrics
are discussed.

\section{Basic equations}

\subsection{Charge transport in graphene}

A graphene layer consists of two triangular sublattices (A and B). The
low-energy band is gapless and electronic states are found near two
(electron and hole) cones. The behavior of charge carriers (electrons and
holes) near the Dirac point is governed by the 2D Dirac equation \cite%
{Slonczewski, Semenoff}:
\begin{equation}  \label{Dirac}
v_{F}\left( \mathbf{\vec{\sigma} \cdot \hat{p}}\right) \Psi =(E-V)\Psi,
\end{equation}%
where $\Psi $ is a two-component spinor $(\Psi _{A},\Psi _{B})^{T}$, the
components of a pseudospin matrix $\vec{\sigma}$ are given by Pauli's
matrices, $\mathbf{\hat{p}}$ is the momentum operator, $v_F$ is the Fermi
velocity, $V(x,y)$ is the potential, and $E$ is the state energy. When the
potential depends on one coordinate, $V=V(x)$, the wave function $\Psi
(x,y)$
can be written as $\Psi (x,y)=e^{ik_{y}y}\psi (x)$, and Eq.~(\ref{Dirac})
can be presented in the dimensionless form
\begin{eqnarray}
{\frac{d\psi _{A}}{d\xi }}-\beta \psi _{A} &=&i[\varepsilon -u(\xi )]\psi
_{B},  \nonumber  \label{Basic} \\
{\frac{d\psi _{B}}{d\xi }}+\beta \psi _{B} &=&i[\varepsilon -u(\xi )]\psi
_{A}.
\end{eqnarray}%
Here $\xi =x/d$, $d$ is the characteristic spatial scale of the potential
variations, $\varepsilon =Ed/\hbar v_{F}$, $u=Vd/\hbar v_{F}$, and $\beta
=k_{y}d$.

\begin{figure}[tbh]
\centering \scalebox{0.3}{\includegraphics{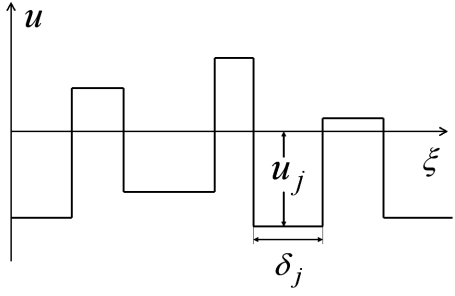}}
\caption{Schematic diagram of the potential $u(\protect\xi)$.}
\label{Fig_1}
\end{figure}

In what follows, we consider potentials $u(\xi )$ comprised of periodic or
random chains of rectangular barriers depicted in Fig.~\ref{Fig_1}
In a $j$-th layer, the solution of Eq.~(\ref{Basic}) has the form: $\psi
_{(A,B)_{j}}=\psi _{(A,B)_{j}}^{(+)}e^{i\kappa _{j}\xi }+\psi
_{(A,B)_{j}}^{(-)}e^{-i\kappa _{j}\xi }$, where $\kappa _{j}=\sqrt{%
(\varepsilon -u_{j})^{2}-\beta ^{2}}$, $\psi _{(A,B)_{j}}^{(\pm )}$ are the
amplitudes of the rightward $(+)$ and leftward $(-)$ propagating spinor
components. At the layer interfaces the amplitudes $\psi _{(A,B)_{j}}^{(\pm
)}$ and $\psi _{(A,B)_{j+1}}^{(\pm )}$ are connected by the equation
\begin{equation}
\psi _{(A,B)_{j+1}}^{(+)}+\psi _{(A,B)_{j+1}}^{(-)}=\psi
_{(A,B)_{j}}^{(+)}+\psi _{(A,B)_{j}}^{(-)},  \label{interface}
\end{equation}%
which follows from the continuity of the spinor components. Since from
Eq.~(%
\ref{Basic}) the amplitudes $\psi _{(A,B)_{j}}^{(\pm )}$ are connected,
\begin{equation}
\psi _{B_{j}}^{(\pm )}=c_{j}^{(\pm )}\psi _{A_{j}}^{(\pm )},\hspace{3mm}%
c_{j}^{(\pm )}={\frac{i\beta \pm \kappa _{j}}{\varepsilon -u_{j}}},
\label{Amplitudes}
\end{equation}%
\begin{equation}
c_{j}^{(\pm )}={\frac{i\beta \pm \kappa }{\sqrt{\beta ^{2}+\kappa ^{2}}}}%
\mathrm{sgn}(\varepsilon -u_{j})=\pm e^{\pm i\theta _{j}}\cdot \mathrm{sgn}%
(\varepsilon -u_{j}),  \label{Amplitudes-a}
\end{equation}%
we will only consider the amplitudes $\psi _{A}^{(\pm )}$ and omit the
subscript ``A''. If $\mathrm{Im}\kappa _{j}=0$, $\theta _{j}=\arctan (\beta
/\kappa _{j})$ is the angle between the wave vector $\mathbf{k}$ of the
$(+)$
wave and the normal to the interface between the $j$-th and $\ (j+1)$-th
layers (i.e., the angle of propagation in the $j$-th layer). Using
Eqs.~(\ref%
{interface}) and (\ref{Amplitudes}) one can calculate the matrix $\hat{M}%
_{j,j+1}$ that connects the amplitudes $\psi _{j}^{(\pm )}$ and $\psi
_{j+1}^{(\pm )}$ on two sides of the interface, $\left( \psi
_{j+1}^{(+)},\psi _{j+1}^{(-)}\right) ^{T}=\hat{M}_{j,j+1}\left( \psi
_{j}^{(+)},\psi _{j}^{(-)}\right) ^{T}$:

\begin{equation}  \label{matrix2}
\hat{M}_{j,j+1}={\frac{1}{2\cos \theta _{j+1}}}\left\Vert
\begin{array}{cc}
g_{j,j+1}^{(+)} & g_{j,j+1}^{(-)} \\
(g_{j,j+1}^{(-)})^{\ast } & (g_{j,j+1}^{(+)})^{\ast }%
\end{array}
\right\Vert ,
\end{equation}
where
\begin{equation}  \label{matrix2-a}
g_{j,j+1}^{(\pm )}=e^{-i\theta _{j+1}}\pm e^{\pm i\theta _{j}}\cdot
\mathrm{%
sgn}[(\varepsilon -u_{j})(\varepsilon -u_{j+1})],
\end{equation}
and $T$ denotes transposed vector.

\begin{figure}[tbh]
\centering \scalebox{0.35}{\includegraphics{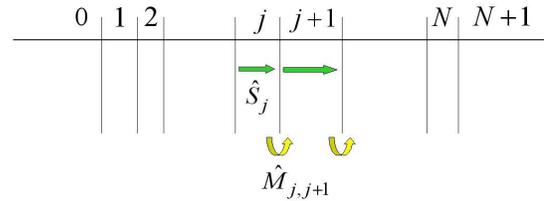}}
\caption{(color online) Propagation through a layered structure. The matrix
$%
\hat{S}_j$ propagates through $j-$th layer (green arrow) and the matrix $%
\hat{M}_{j,j+1}$ connects the spinor amplitudes across the interface between
$j$-th and $(j+1)$-th layers.}
\label{Fig_2}
\end{figure}

The spinor components at the left and right boundaries of the $j$-th layer
are connected by the diagonal matrix $\hat{S}_{j}=\mathrm{diag}%
(e^{i\alpha_{j}},e^{-i\alpha _{j}})$, where $\alpha_{j}=\kappa _{j}\delta
_{j}$ is the phase accumulated by the wave propagating through the layer of
the thickness $\delta _{j}$. Thus, the matrix $\hat{S}_{j+1}\hat{M}_{j,j+1}$
transports the spinor components from the left side of the interface between
the $j$-th and $(j+1)$-th layers to the left side of the next interface
between the $(j+1)$-th and $(j+2)$-th layers (see Fig.~\ref{Fig_2}).
Obviously, the total transfer matrix of a layered sample consisting of $N$
layers is given by the product of $\hat{S}\hat{M}$ matrices:

\begin{equation}  \label{product}
\hat{T}=\prod\limits_{j=0}^{N}\hat{S}_{j+1}\hat{M}_{j,j+1}
\end{equation}

\subsection{Light transport in dielectrics}

Analogous products of matrices have been well studied in the context of
transport of electromagnetic waves in layered media (see, for example,
\cite%
{matrices} and references therein). To better understand the
physics of charge transport in graphene subject to a
coordinate-dependent potential, in what follows, we contrast the
results for graphene with those for the propagation of light in
layered dielectric media (for more analogies between quantum and
optical systems, see, e.g. Ref.\cite{Dragoman, analogy}).
Additional analogies, not discussed here, also exist with the
transport and localization of phonons in different kinds of
periodic and random one-dimensional structures \cite{Nori, Nori_1,
Nori_2}.

In the latter case, the matrices $\hat{S}_{j}$ are the same as in Eq.~(\ref%
{product}), and the transfer matrix, $\hat{\mathcal{M}}_{j,j+1}$,
\begin{equation}
\hat{\mathcal{M}}_{j,j+1}={\frac{1}{2\cos \theta _{j+1}}}\left\|
\begin{array}{cc}
\mathcal{G}_{j,j+1}^{(+)} & \mathcal{G}_{j,j+1}^{(-)} \\
(\mathcal{G}_{j,j+1}^{(-)})^{\ast } & (\mathcal{G}_{j,j+1}^{(+)})^{\ast }%
\end{array}%
\right\| ,  \label{matrix3}
\end{equation}%
that describes the transformation of the amplitudes of the electromagnetic
waves at the interface between $j$th and $\ (j+1)$th layers, has the form
Eq.~(\ref{matrix2}) with $g_{j,j+1}^{(\pm )}$ being replaced by
\begin{equation}
\mathcal{G}_{j,j+1}^{(\pm )}=\cos \theta _{j+1}\pm \cos \theta _{j}\cdot
\mathrm{sgn}(n_{j}n_{j+1}){\frac{Z_{j+1}}{Z_{j}}}  \label{RLmatrix-1}
\end{equation}%
for $s$-polarized waves and
\begin{equation}
\mathcal{G}_{j,j+1}^{(\pm )}={\frac{Z_{j+1}}{Z_{j}}}\cos \theta _{j+1}\pm
\cos \theta _{j}\cdot \mathrm{sgn}(n_{j}n_{j+1})  \label{RLmatrix-2}
\end{equation}%
for $p$-polarized waves.\ Here, $\theta _{j}$ is the angle of the
propagation, $Z_{j}=\sqrt{\mu _{j}/\varepsilon _{j}}$ is the impedance of $\
j$th layer, and $n_{j}=\pm \sqrt{\varepsilon _{j}\mu _{j}}$ is its
refractive index. The signs $\pm $ correspond, respectively, to dielectrics
with positive (right-handed, R) and negative (left-handed, L) refractive
indices.

It is easy to see that the parameter $(\varepsilon -u)$ plays, in graphene,
the same role as the refractive index $n$ in a dielectric medium. It is due
to this similarity that a p-n junction (interface between regions where the
values $(\varepsilon -u)$ have opposite signs) focuses charge carriers in
graphene, like an R-L interface focuses electromagnetic waves
\cite{Cheianov}%
.

Note that in Eq.~(\ref{matrix2-a}) (for graphene) there is no factor $%
Z_{j+1}/Z_{j}$, which determines the reflection coefficients at the boundary
between two dielectrics \cite{born wolf}. This means that the charge
transport in graphene is similar to the propagation of light in a stack of
dielectric layers with equal impedances. In particular, both p-n and p-p
junctions are transparent for normally incident charged particles \cite%
{Cheianov, Katsnelson1}. This property is readily seen from the analysis of
the matrix $\hat{M}$: at $\beta =0$ it is a Pauli matrix $\hat{M}=\sigma_x$
for p-n junctions and unit matrix for p-p junctions. Therefore, a $(+)$-wave
is totally transformed into a $(-)$-wave at a p-n junction, and remains a $%
(+)$-wave at a p-p junction. Another important difference between the
transfer matrices $\hat{M}$ (graphene) and $\hat{\mathcal{M}}$
(electromagnetic waves) is that $\hat{M}$ is, generally speaking, a
complex-valued matrix, while the $\hat{\mathcal{M}}$ is always real. As it
is shown bellow, this distinction brings about rather peculiar
dissimilarities between the conductivity of graphene and the transparency of
dielectrics.

\section{Transport in periodic structures}

Among the vast amount of publications on graphene, a significant and ever
increasing part belongs to papers devoted to the charge transport in
graphene superlattices formed by a periodic external potential (see, e.g.,
\cite{periodic-1, periodic-2, periodic-3, periodic-4,periodic-5}). This is
not only due to its theoretical interest but also because of the possibility
of experimental realization and potential applications \cite{periodic-4}. In
Ref. \cite{periodic-1}, for example, it was suggested that, by virtue of the
high anisotropy of the propagation of carriers through graphene subjected to
a Kronig--Penney type periodic potential, such a structure could be used for
building graphene electronic circuits from appropriately engineered periodic
surface patterns.

Here we consider a layer of graphene under a periodic alternating potential
$%
u=\pm U_{0}$ and assume that $\varepsilon =0$. Two layers, $u_{1}=U_{0}$ and
$u_{2}=-U_{0}$ with thicknesses $\delta _{1}$ and $\delta _{2}$,
respectively, constitute the superlatice period with transfer matrix
$\hat{T}%
=\hat{S}_{2}\hat{M}_{2,1}\hat{S}_{1}\hat{M}_{1,2}.$ Its eigenvalues $\lambda
(\beta )$ indicate whether this periodic structure is transparent or not.
Namely, the structure is transparent when $|\lambda (\beta )|$ $=1\ $and is
opaque when $|\lambda (\beta )|$ $\neq 1$. Note that analogous ($%
n_{1}=-n_{2} $) L-R periodic dielectric structures are transparent at all
angles of incidence when $Z_{1}=Z_{2}.$ In contrast, a periodic array of p-n
junctions in graphene has a rather nontrivial angular dependence of the
transmission coefficient $T(\beta )$. This distinction between periodic
graphene and dielectric lattices follows from the difference in the
corresponding transfer matrices: $\hat{M}$ is complex-valued while $\hat{%
\mathcal{M}}$ is real.

\begin{figure}[tbh]
\centering \scalebox{0.5}{\includegraphics{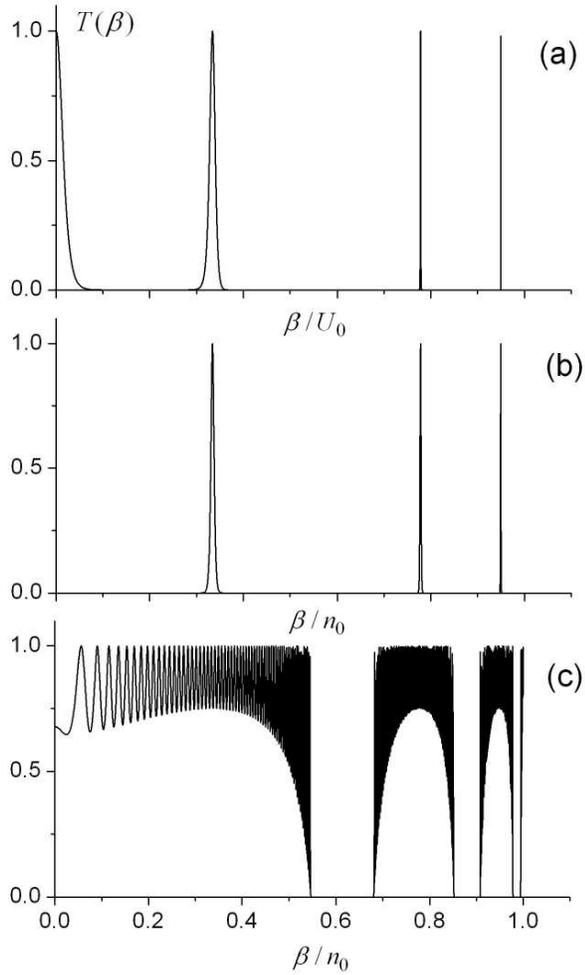}}
\caption{Transmission coefficient $T(\protect\beta)$ for: (a) graphene
subject to a symmetric periodic potential $u=\pm U_0$; (b) a symmetric
periodic L-R dielectric structure; (c) a symmetric periodic R-R dielectric
structure. The parameter $U_0$ plays in graphene the same role as the
refractive index $n_{0}$ in a dielectric medium. For (b) and (c), $%
n_{0}=U_{0}$ and $Z_{1}/Z_{2}=1.1$. In all figures, L stands for left-handed
media and R for right-handed media. Thus, the structures considered in (b)
are periodic stacks LRLRLR\ldots of dielectrics. The structures considered
in (c) are periodic RRRRRR\ldots stacks of dielectrics.}
\label{Fig_3}
\end{figure}

If $\varepsilon =0$ and $\delta _{1}=\delta _{2}\equiv \delta $ (symmetric
graphene system) the equation for the eigenvalues of the matrix $\hat{T}$
has the form
\begin{equation}
\lambda ^{2}-2\lambda {\frac{1-\sin ^{2}\theta \cos 2\alpha }{\cos
^{2}\theta }}+1=0,
\end{equation}%
where $\alpha =\kappa (\theta )\delta $. It is easy to see that $|\lambda
|=1 $ at normal incidence $(\theta=0)$ and at a discrete set of angles, $%
\theta _{m},$ given by
\begin{equation}
\alpha (\theta _{m})=m\pi ,\hspace{3mm}m=\pm 1,\pm 2,\ldots
\label{resonance}
\end{equation}%
Although the eigenvalues $\lambda (\beta )$\ are well defined for an
infinite system, they are also quite meaningful for a sufficiently long
finite periodic sample: at $\theta _{m}$\ there are maxima of the
transmission coefficient $T(\beta ).$ The transmission coefficient $T(\beta
) $ of the symmetric graphene structure is presented in Fig.~\ref{Fig_3}a. a
similar transmission spectrum $T(\beta )$ exists at $\theta \neq 0$ in L-R
periodic structures (Fig.~\ref{Fig_3}b) made of layers with equal absolute
values of the refractive indexes, $n_{1}=-n_{2}$, and different impedances
$%
Z_{1}\neq $ $Z_{2}$ \cite{Wu1,Wu2}. The transmission coefficient $T(\beta )$
for a symmetric periodic R-R system is shown in Fig.~\ref{Fig_3}c.

\begin{figure}[tbh]
\centering \scalebox{0.55}{\includegraphics{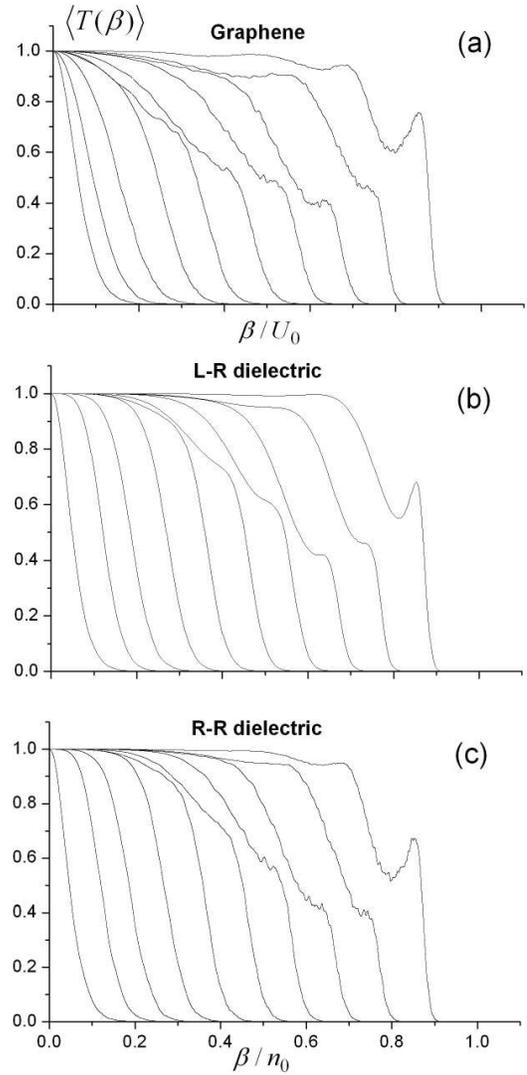}}
\caption{(color online) Transmission coefficient $T(\protect\beta,\Delta)$
as a function of the difference of the thicknesses of two layers, $\Delta=(%
\protect\delta_1-\protect\delta_2)/2$, and of the normalized parameter $%
\protect\beta=k_{y}d\equiv kd \sin \protect\theta$ for: (a) periodic
graphene subject to an alternating periodic potential; (b) periodic L-R
dielectric structure; (c) periodic R-R dielectric structure. In (a), the
normalization parameter $U_{0}$ is the dimensionless applied voltage; in (b)
and (c), the normalization parameter $n_{0}$ is the refraction index in a
dielectric medium, and $Z_1/Z_2=1.1$. The grey area corresponds to a perfect
transmission, $T = 1$. The white regions correspond to $T = 0$.}
\label{Fig_4abc}
\end{figure}

The transmission spectrum $T(\beta )$ in Fig.~\ref{Fig_3}a, which consists
of a discrete set of incidence angles, is the result of the degeneracy
caused by the high symmetry of the structure ($u_{1}=-u_{2}$, $\varepsilon
=0 $, and $\delta _{1}=\delta _{2}$). Any symmetry-breaking splits the
degeneracy and the spectrum takes the form usual for ordinary periodic
structures: a set of conducting zones of non-zero width separated by band
gaps \cite{periodic-2}. In Fig.~\ref{Fig_4abc}a, the zone structure of the
transmission spectrum is shown. It is important to note that instead of the
wave number $k_{y}$ and the energy, typically used in zone diagrams, the
variables in Fig.~\ref{Fig_4abc}a are the asymmetry parameter $\Delta
=(\delta _{1}-\delta _{2})/2$\ and $\beta =kd\sin \theta ,$ respectively.
Note that even in non-symmetric structures there are some values of $\delta
_{1}$ and $\delta _{2}$ for which, along with the usual conducting zones and
band gaps, there exists a discrete set of resonant $\beta $s. Note also
that, for a fixed $\Delta $, the transmission zones as a function of $\beta
$
are very narrow, making the direction of the charge flux easily tunable by
changing the applied voltage $U_{0}$.

For comparison, the analogous spectra for L-R and R-R periodic structures
with $|n_{1}|=|n_{2}|$ and $Z_{1}\neq Z_{2}$ are presented in Fig.~\ref%
{Fig_4abc}b,c. One can see there that the transmission coefficients of
graphene and L-R structures are similar, and both differ drastically from
that of the R-R structure.

A phenomenon similar to the total internal reflection of light can occur to
charge in graphene at a non-symmetric p-n interface when $%
|\varepsilon-u_{1}|\neq |\varepsilon -u_{2}|$ (here $u_{1}$ and $u_{2}$ are
the potentials on either side of the interface). However, a periodic set of
such junctions is transparent for some angles of incidence
\cite{periodic-2}%
. This effect is similar to photon tunneling (frustrated total internal
reflection) in stacks of dielectric layers \cite{Dragila}.

\begin{figure}[tbh]
\centering \scalebox{0.55}{\includegraphics{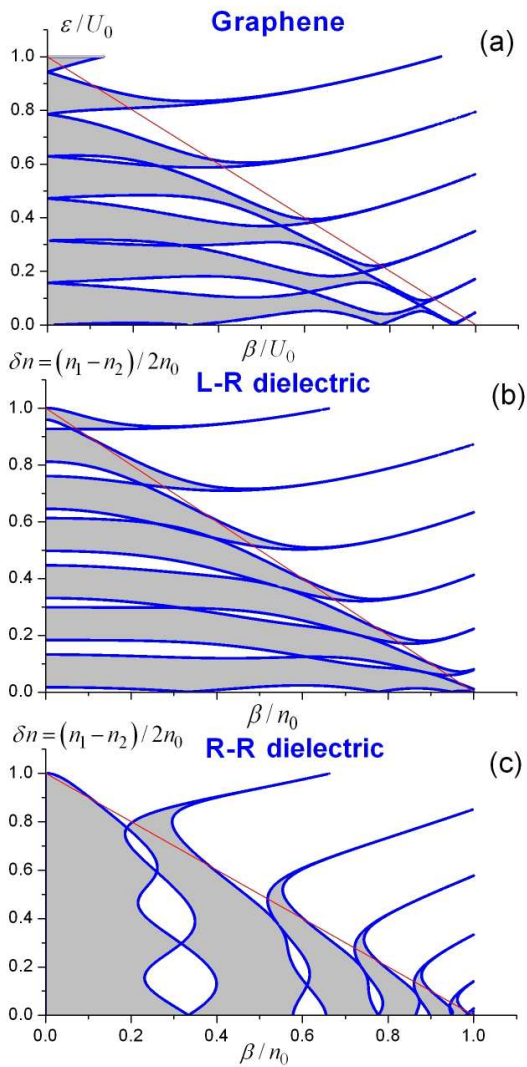}}
\caption{(color online) Transmission coefficient $T(\protect\beta)$ for: (a)
periodic graphene subject to an alternating potential; (b) periodic L-R
dielectric structure; (c) periodic R-R dielectric structure. The coordinates
and the colors are the same as in Fig.~\protect\ref{Fig_4abc}. In the region
above the straight line the condition of total internal reflection is
satisfied.}
\label{Fig_5abc}
\end{figure}

Let us now consider the transmission of charge through a single
non-symmetric p-n junction, assuming that $u_{1}=-u_{2}=U_0\geq\epsilon>0$.
Then, if $(\varepsilon-U_0)^2<\beta^2<(\varepsilon+U_0)^2$, a total internal
reflection occurs at the interface; however there are ranges of the angle of
incidence where the system is transparent. An example of such an angular
spectrum is shown in Fig.~\ref{Fig_5abc}a.

Physically, the tunneling conductance (proportional to the transmission
coefficient) is due to the confined states in graphene quantum well \cite%
{Pereira, RIKEN}. Although the confined states in a single quantum well have
a discrete spectrum $\beta =\beta _{n}(\varepsilon )$, an infinite periodic
chain of wells, interacting via their evanescent wave functions, forms
transmission bands centered around $\beta _{n}$.

Periodic L-R and R-R dielectric structures have properties similar to the
properties described in this subsection for graphene. Figures
\ref{Fig_5abc}%
b,c show the transmission spectrum of L-R and R-R structures with a period
composed of two blocks with equal thicknesses and different refractive
indexes, $|n_1|=1+\delta n$ and $|n_2|=1-\delta n$.

\section{Transmission in disordered structures}

Based on the results obtained for ideally periodic systems, the authors of
Ref.~\onlinecite{periodic-1, periodic-4} suggested that graphene
superlattices could be used as tunable elements in electronic devices. Since
parameters of such structures are extremely sensitive to the variations of
the applied potential it is worthwhile to study the effect of disorder
(random deviations of the potential from periodicity) on the propagation of
charge in such configurations. Moreover, this study is of interest by itself
because strongly disordered (with no periodic component) potentials bring
about further unexpected spectral and transport properties of graphene
samples, which make them potentially useful as an alternative to pure
periodic systems.

A surprising and counter-intuitive result is that a sample of graphene
subject to a random one-dimensional potential, $u(x),$ is absolutely
transparent to the charge flow perpendicular to the $x$-direction, no matter
how long is the sample and how strong the disorder is\cite{Titov-1}. This
means that in such samples there exists a minimal non-zero conductivity,
which (together with symmetry and spectral flow arguments) led to the
conclusion that there is no localization in 1D disordered graphene systems
\cite{Titov-1, Koshino}. However, this statement (being correct in some
sense) should be perceived with a certain caution. Below, we show that
although the wave functions of normally incident particles are extended and
belong to the continuous part of the spectrum, away from some vicinity of $%
\theta =0,$ \textit{1-D random graphene systems manifest all features of
disorder-induced strong localization}. Indeed, there exist a discrete random
set of angles (or a discrete random set of energies for each given angle)
for which the corresponding wave functions are exponentially localized with
a Lyapunov exponent (inverse localization length) proportional to the
strength of the disorder.

Obviously, the behavior of a quantum mechanical particle is determined by
the type of potential and by the ratio between its values for $u(x)$ and the
energy $\varepsilon $ of the particle. Bellow, we study the charge transport
in graphene subject to a random layered potential of the form $%
u_{j}=u_{0}(j)+\Delta u_{j}.$ Three particular cases are considered: (i) all
$u_{j}<\varepsilon $, $u_{0}(j)$ is a periodic function; (ii) $\varepsilon
\leq u_{0}(j)=\mathrm{const}$; and (iii) $\varepsilon =0$, $u_{0}(j)$ is a
periodic set of numbers with alternating signs. In all cases, the $\Delta
u_{j}$ are independent random variables homogeneously distributed in the
interval $[-\Delta u_{0},\Delta u_{0}]$. In (iii), $\left| \Delta
u_{j}\right| <\left| u_{0}(j)\right| $ and $u_{0}(j)=\pm U_{0}$, which
represents an array of random p-n junctions, where electrons outside a
barrier transform into holes inside it, or vice versa. For the sake of
simplicity, here we assume that the widths of the layers do not fluctuate.
%The geometrical disorder (fluctuating width $\delta _{j}$ of the layers)
will be addressed latter. These three cases will be considered in the next
three subsections.

\subsection{Case (i): all $u_j<\protect\varepsilon$, and $u_0(j)$ periodic}

\begin{figure}[tbh]
\centering \scalebox{0.4}{\includegraphics{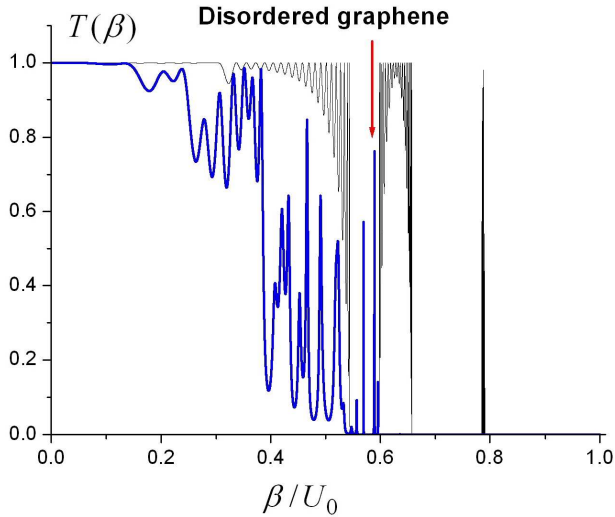}}
\caption{(color online) Transmission coefficient $T(\protect\beta)$ for
periodic (thin black line) and disordered (bold blue line) graphene. The
range of the variation of the potentials, $\Delta u_0=0.1U_{0}$; $\protect%
\varepsilon=0$.}
\label{Fig_6}
\end{figure}

\begin{figure}[tbh]
\centering \scalebox{0.4}{\includegraphics{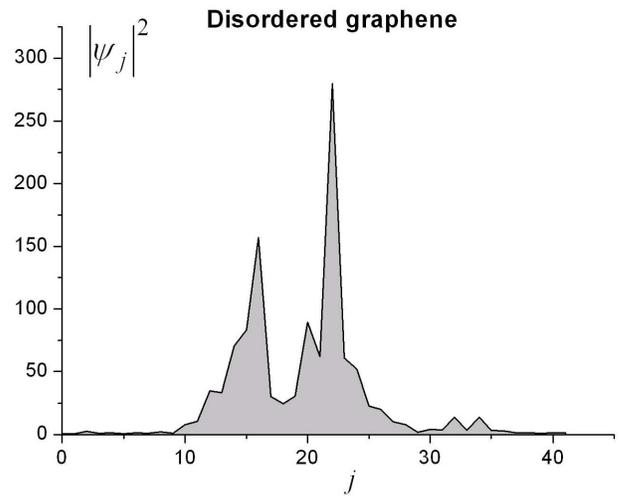}}
\caption{Spatial distribution of the wave function localized inside the
sample for $\protect\beta$ marked by a red arrow in
Fig.~\protect\ref{Fig_6}.
}
\label{Fig_7}
\end{figure}

Figure Fig.~\ref{Fig_6} shows an example of the angular dependence of the
transmission coefficient, $T(\beta )$, for a type (i) graphene sample that
contains $40$ layers of equal thickness $\delta _{0}=1.0$, $u_{1}=-7.0$, $%
u_{2}=-13.0$, $\varepsilon =0.0$.
One can see that a relatively weak disorder has drastically changed the
transmission spectrum: all features of the spectrum of the underlying
periodic structure has been washed out, and a rather dense (quasi-)discrete
angular spectrum has appeared with the corresponding wave functions
localized at random points inside the sample (disorder-induced resonances).
Figure~\ref{Fig_7} shows the spatial distribution of the square modulus of
the amplitude of a resonant wave function (intensity distribution inside the
sample).
For a fixed $\varepsilon $, $T(\beta )$, shown in Fig.~\ref{Fig_6}, has the
same form as $T(\varepsilon )$ for a fixed $\beta $, shown in Fig.~\ref%
{Fig_8}. Both consist of randomly distributed resonances (one in the $\beta
$
domain and another versus energy) typical for 1D Anderson localization of
electrons and light. However, there is one fundamental difference from the
usual Anderson localization: in the vicinity of normal incidence, the
transmission spectrum of graphene is continuous with extended wave
functions, and the transmission coefficient is finite ($T=1$ at $\theta
=0)$%
. It is this range of angles that provides the finite minimal conductivity,
which is proportional to the integral of $T(\theta )$ over all angles $%
\theta $.

\begin{figure}[tbh]
\centering \scalebox{0.4}{\includegraphics{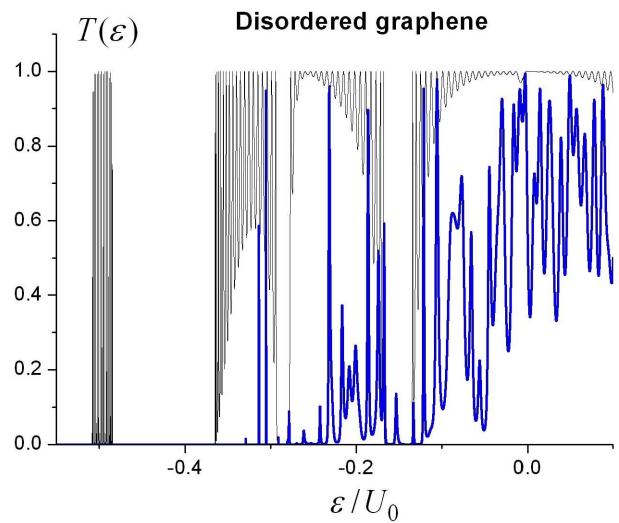}}
\caption{(color online) Transmission coefficient as a function of the energy
$\protect\varepsilon$, $T(\protect\varepsilon)$, for periodic (thin black
line) and disordered (bold blue line) graphene. $\Delta u_{0}=0.1U_{0}$, $%
\protect\beta=0.3U_{0}$.}
\label{Fig_8}
\end{figure}

\begin{figure}[tbh]
\centering \scalebox{0.4}{\includegraphics{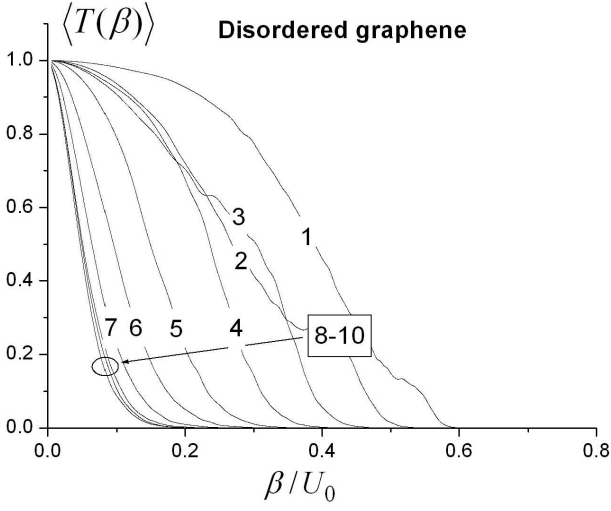}}
\caption{Mean transmission coefficient $\left\langle T(\protect\beta%
)\right\rangle$ for disordered graphene. The curve (1) corresponds to $%
\Delta u_{0}/U_{0}=0.1$, the curve (2) corresponds to $\Delta
u_{0}/U_{0}=0.2
$, etc. Curves (8), (9), and (10), for which $\Delta u_{0}/U_{0}=0.8$, 0.9,
and 1.0 are practically indiscernible.}
\label{Fig_9}
\end{figure}

The mean transmission coefficient, $\langle T(\beta )\rangle $, for
different strengths of disorder (different $\Delta u_{0}$) is plotted in
Fig.~\ref{Fig_9}. As expected, the increase of disorder reduces the
transmission and narrows down the angular width of the transmission spectrum
$\Delta \beta $. The zero (to within the resolution of the plots) values of
$%
\langle T(\beta )\rangle $ at each curve correspond to the angles{\LARGE ,}
which exceed the angle of total internal reflection.

\subsection{Case (ii): $\protect\varepsilon\leq u_0(j)=\mathrm{const}$}

\begin{figure}[tbh]
\centering \scalebox{0.4}{\includegraphics{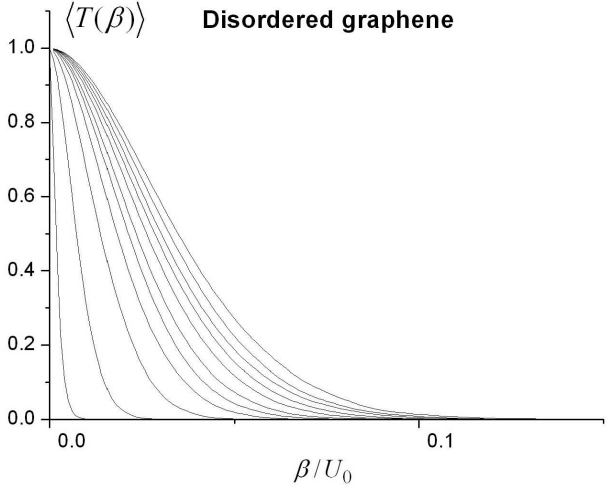}}
\caption{Mean transmission coefficient $\left\langle T(\protect\beta%
)\right\rangle$ for disordered graphene for different values of the
parameter $\Delta u_{0}/U_{0}$, which increases from zero (narrowest curve)
to 1.0 (widest curve). Here, $\Delta u_0/U_0$ increases by 0.1 from left to
right.}
\label{Fig_10}
\end{figure}

In this case, the results are more intriguing \cite{Titov-1,San-Jose}
(although encountered in usual electron and optical random systems \cite%
{Fr-enhancement}). In this case, the transmission of the unperturbed system
is exponentially small (tunneling) and gets enhanced by the fluctuation of
the potential (Fig.~\ref{Fig_10}).
This is quite natural because the transmission of each $j$-th segment is
proportional to $\exp [-\delta _{0}(\varepsilon \pm \Delta u_{j})]$ and
(despite of the fact that $\langle \Delta u_{j}\rangle =0$) the mean value
$%
\left\langle \exp [-\delta _{0}(\varepsilon \pm \Delta u_{j})]\right\rangle
>\exp (-\delta _{0}\varepsilon )$, due to the asymmetry of the exponential
function. Note that another type of disorder, linked to graphene
layer edges, leads to the same result: the disorder improves the
transmission \cite{Rozhkov}, compared to the ordered graphene
case.

\subsection{Case (iii): $|\Delta u_j|<|u_0(j)|$ and $u_0(j)=\pm U_0$}

\begin{figure}[tbh]
\centering \scalebox{0.4}{\includegraphics{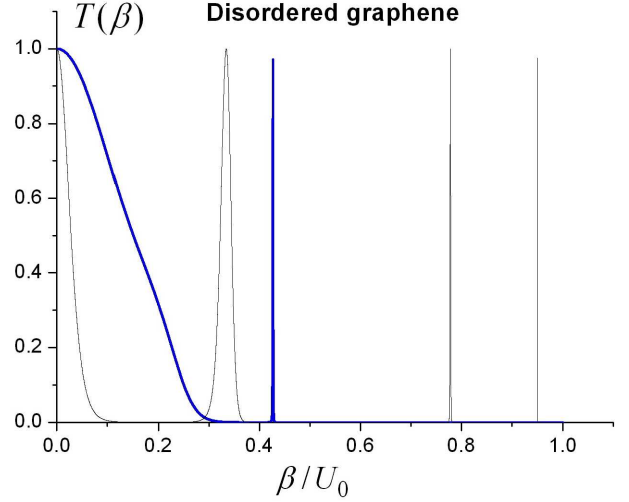}}
\caption{(color online) Transmission coefficient $T(\protect\beta)$ for
periodic (thin black line) and random (bold blue line) symmetric potential
with $\Delta u_{j}\in (-0.1, 0.1)$.}
\label{Fig_11}
\end{figure}

The behavior of the charge carriers in the graphene system of type (iii) is
most unusual. It is characteristic of two-dimensional Fermions and have no
analogies in electron and light transport. Shown in Fig.~\ref{Fig_11} by the
bold blue line is the transmission spectrum at $\varepsilon =0$ of a
graphene sample containing 40 layers of equal thicknesses $\delta
_{j}=\delta _{0}$ and alternating random potential. One can see that,
compared to the underlying periodic configuration (thin black line), the
disorder: obliterates the transmission peaks located near $\beta _{m}$ with
$%
m\neq 1$ [see Eq.~(\ref{resonance})]; makes much wider the transparency zone
near $\beta =0$; {\small and }gives rise to a new narrow peak in the
transmission coefficient, associated with wave localization in the random
potential. In contrast to the peaks in the periodic structure, the wave
function of this disorder-induced resonance is exponentially localized.
\begin{figure}[tbh]
\centering \scalebox{0.4}{\includegraphics{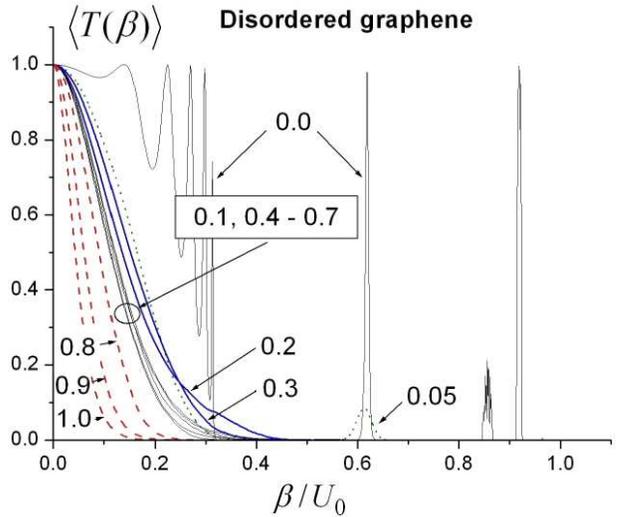}}
\caption{(color online) Mean transmission coefficient $\left\langle T(%
\protect\beta)\right\rangle$ for disordered graphene. Arrows with numbers
mark the strength of the disorder, $\Delta u_{0}/U_{0}$.}
\label{Fig_12}
\end{figure}

\begin{figure}[tbh]
\centering \scalebox{0.4}{\includegraphics{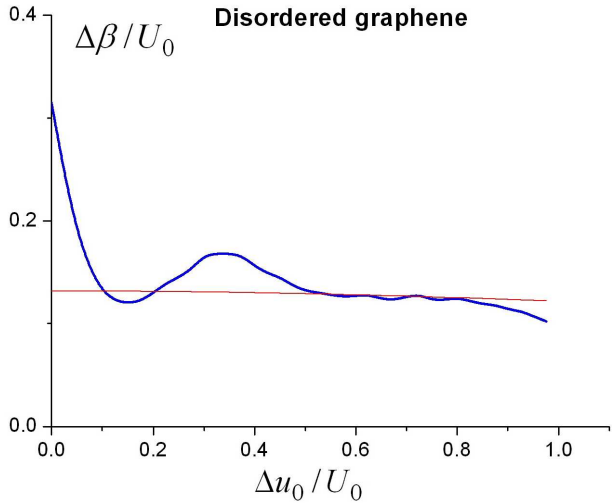}}
\caption{(color online) Half-width $\Delta\beta$ of the angular spectrum for a (iii)
structure. Red thin line corresponds to the half-width $\Delta\beta$
given by Eq.~(\protect\ref{Eq.105}).}
\label{Fig_13}
\end{figure}

For this case (iii), the average transmission coefficient as a function of
$%
\beta $ is presented in Fig.~\ref{Fig_12}. In contrast to the case (i), the
transmission in (iii) is extremely sensitive to fluctuations of the applied
potential: in Fig.~\ref{Fig_12} the relative fluctuations $\Delta
u_{0}/u_{0}=0.05$ reduce the angular width of the transmission spectrum more
than four times (see also Fig.~\ref{Fig_13}).
That high sensitivity makes such a system a good candidate for use in
electronic circuits capable of tuning the direction of charge flow. Another
striking property is that after this abrupt drop in the transmission, it
(i.e., the transmission) becomes practically independent of the strength of
disorder in a relatively large range, as shown in Fig.~\ref{Fig_13}.

\begin{figure}[tbh]
\centering \scalebox{0.6}{\includegraphics{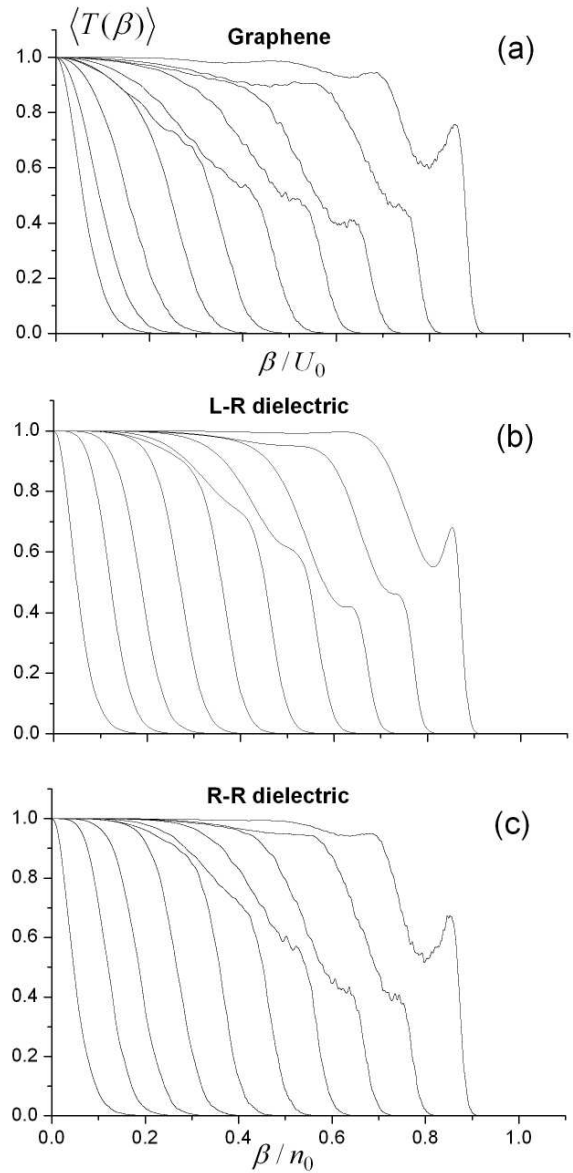}}
\caption{Mean transmission coefficient $\left\langle T(\protect\beta %
)\right\rangle $ for: (a) disordered type (ii) graphene structure; (b)
disordered L-R dielectric structure; (c) disordered R-R dielectric
structure. Different curves correspond to the different values of the
strength of the disorder $\Delta u_{0}/U_{0}$ in (a), $\protect\delta
n/n_{0}
$ in (b) and (c), which increases from right to left from 0.1 to 1.0 with a
step of 0.1.}
\label{Fig_14abc}
\end{figure}

The propagation of light in analogous L-R and R-R disordered dielectric
structures demonstrates completely different behavior. As the degree of
disorder (variations $\Delta n_{j}$ of the refractive indexes $n_{j})$
grows, the averaged angular spectra quickly reach their asymptotic
\textquotedblleft rectangular\textquotedblright\ shape: a constant
transmission in the region where all interfaces between layers are
transparent followed by an abrupt decrease in transmission in the region of
$%
\beta $ where the total internal reflection appears (see
Fig.~\ref{Fig_14abc}%
b,c). The frequency dependences of the transmission coefficient and
localization length have been studied in Ref.~\onlinecite{Asatryan}.

\section{Analytical study}

The features presented above can be explained, both qualitatively and
quantitatively, in the framework of a rather simple theoretical approach. It
can be shown \cite{matrices} that in the short-wavelength limit, $k\delta
\gg1,$ the mean amplitude transmission coefficient, $\left\langle
T^{(N)}\right\rangle$, of a sample built of $N$ layers, is approximately
equal to
\begin{equation}  \label{Eq 101}
\left\langle T^{(N)}\right\rangle\simeq{\prod}_{j=1}^{N}\left%
\langle|t_{j,j+1}|^{2}\right\rangle,
\end{equation}%
where $t_{j,j+1}$ are statistically independent complex transmission
coefficients of the boundaries between the $j$-th and $(j+1)$-th layers, and
\begin{equation}  \label{Eq. 102}
|t_{j,j+1}|^{2}={\frac{2\cos ^{2}\theta _{j+1}}{1+\cos (\theta
_{j}-\theta_{j+1})},}
\end{equation}%
At small $\theta \ll1$, equation (\ref{Eq. 102}) becomes
\begin{equation}  \label{Eq. 102-aa}
\left\vert t_{j,j+1}\right\vert ^{2} \simeq 1-{\frac{3}{4}}\theta
_{j+1}^{2}+%
{\frac{1}{4}}\theta _{j}^{2}-{\frac{1}{2}}\theta _{j}\theta_{j+1},
\end{equation}
\[
\theta _{j} =\arctan\left[ {\frac{\beta}{\sqrt{(u_{j}-\varepsilon
)^{2}-\beta^{2}}}}\right]\simeq {\frac{\beta }{|u_{j}-\varepsilon |}},
\]
and from Eqs.~(\ref{Eq 101}) and (\ref{Eq. 102-aa}) it follows that
\begin{equation}  \label{Eq. 103}
\left\langle T^{(N)}\right\rangle\simeq\left[1-{\frac{1}{2}}\langle \theta
^{2}\rangle -{\frac{1}{2}}\langle \theta \rangle ^{2}\right]^{N}.
\end{equation}%
In an initially periodic array of alternating p-n and n-p junctions, $%
u_{j}=-u_{j+1}=U_{0},$ at $\varepsilon =0$ [structure (iii)], Eq.~(\ref{Eq.
103}) yields%
\begin{equation}  \label{Eq.104}
\left\langle T^{(N)}(\beta )\right\rangle \simeq 1-{\frac{1}{2}}{\frac{\beta
^{2}}{(\Delta \beta )^{2}}},
\end{equation}%
where
\begin{equation}
\Delta \beta ={\frac{U_{0}}{\sqrt{N}}}\left[{\frac{\ln 2}{1+\langle \delta
u\rangle ^2/2U_0^2}}\right] ^{1/2}  \label{Eq.105}
\end{equation}%
is the half-width of the angular spectrum, defined as the value of $\beta $
where $\langle T^{(N)}(\beta )\rangle =1/2$. Equations (\ref{Eq.104}),
(\ref%
{Eq.105}) fit well the numerical results presented in Figs.~\ref{Fig_12}, %
\ref{Fig_13}. In particular, they describe the numerically observed
quadratic dependence on $\beta $ and surprisingly weak dependence of the
mean transmission on the strength of the disorder.

In a disordered graphene superlattice consisting only of n-n and p-p
junctions [structure (ii)], the mean transmission coefficient $\left\langle
T^{(N)}\right\rangle $ at $\langle \delta u^{2}\rangle /U_{0}^{2}\ll 1$ is
given by
\begin{equation}
\left\langle T^{(N)}\right\rangle \simeq\left[ 1-{\frac{1}{2}}\langle \theta
^{2}\rangle +{\frac{1}{2}}\langle \theta \rangle ^{2}\right] ^{N}\simeq %
\left[ 1-{\frac{1}{2}}{\frac{\beta ^{2}}{U_{0}^{2}}}{\frac{\langle \delta
u^{2}\rangle }{U_{0}^{2}}}\right] ^{N}.  \label{Eq. 106}
\end{equation}%
In this case, it is easy to see that the half-width of the angular spectrum
strongly depends on the strength of the fluctuations and decreases with
increasing $\langle \delta u^{2}\rangle /U_{0}^{2}$, as can be seen in
Fig.~%
\ref{Fig_14abc}a and in the relation:%
\begin{equation}
\Delta \beta ={\frac{U_{0}}{\sqrt{N}}}\left[ {\frac{2\ln 2U_{0}^{2}}{\langle
\delta u\rangle ^{2}}}\right] ^{1/2}.  \label{Eq. 107}
\end{equation}%
Note, that in the cases (i) and (iii), the transmittance spectrum has
parabolic shape for small angles of incidence, $\left\langle T^{(N)}(\beta
)\right\rangle \simeq 1-\beta ^{2}/\Delta \beta ^{2}$, and the spectrum
half-width $\Delta \beta $ decreases as $1/\sqrt{N}$ when the number $N$ of
layers (the sample length $L$) increases. The same spectrum property for
case (ii) has been predicted in Ref.~\onlinecite{San-Jose}.

In contrast to the charge transport in disordered graphene superlattices
described above, the propagation of light in randomly layered dielectrics is
similar (at $\theta \ll1$) for L-R and RR arrays of layers with equal
impedances (there are the analogs of p-n and p-p junctions, respectively).
This follows from the fact that in both cases the small-angle asymptotics of
the mean transmission coefficient through a boundary between layers are
identical and at $\theta \ll1$ have an universal form (compare with Eqs.
\ref%
{Eq. 103}, \ref{Eq.104}):
\[
\left\langle \left| t_{j,j+1}\right| ^{2}\right\rangle \simeq 1-\frac{1}{2}
\left\langle \theta_{j+1}^{2}\right\rangle+\frac{1}{2} \left\langle
\theta_{j}^{2}\right\rangle,
\]%
which yeilds
\begin{equation}
\left\langle T^{(N)}\right\rangle \simeq 1-O\left( \theta ^{4}\right) .
\label{eq108}
\end{equation}

As in periodic systems, the difference in the transmission spectra of
disordered graphene and dielectric samples [compare Eqs.~(\ref{Eq 101}), (%
\ref{Eq. 106}), and (\ref{eq108})] is a consequence of the above mentioned
absence of imaginary part in the transfer matrix $\hat{\mathcal{M}}$, Eq.~(%
\ref{matrix3}). Examples of the numerically calculated (with no
approximations) angular spectra of the transmission of light are shown in
Fig.~\ref{Fig_14abc}b,c.

\section{Geometrical disorder}

\begin{figure}[tbh]
\centering \scalebox{0.4}{\includegraphics{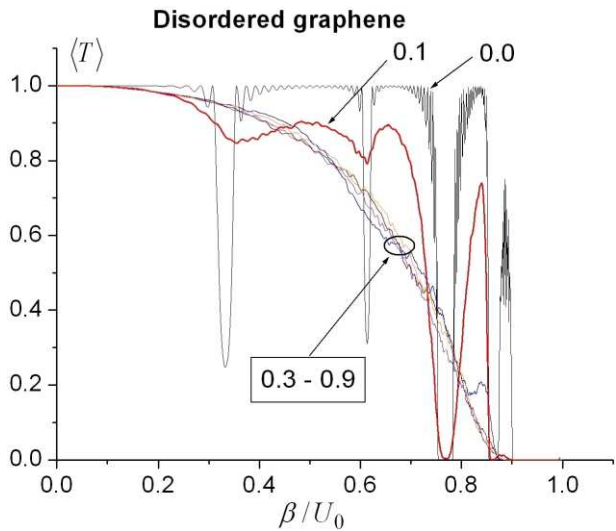}}
\caption{(color online) Mean transmission coefficient $\left\langle T(%
\protect\beta)\right\rangle$ for graphene with geometrical disorder (section
VI). Arrows with numbers mark the strength of disorder, $\Delta/\protect%
\delta$.}
\label{Fig_15}
\end{figure}

In Section IV, we studied spatially-periodic layered graphene structures, in
which the values of the applied potential in each layer were statistically
independent random numbers. Further numerical calculations show that the
main features of the transport and localization of charge in disordered
graphene superlatices are rather universal, i.e., independent of the type of
disorder. When instead of the amplitude of the potential, the size of each
layer is fluctuating,
\[
\Delta u_{j}=0,\hspace{3mm}\delta_{j}=\delta+\Delta_j
\]
all results are similar, at least qualitatively. In Fig. \ref{Fig_15}, the
angular dependences of the mean transmission coefficient, $\left\langle
T(\beta)\right\rangle$, are plotted for different strengths of the
geometrical disorder (different values of $\Delta/\delta$) in the case when
$%
u_j<\varepsilon $ and assuming that $\Delta_j$ are independent random and
homogeneously distributed in the interval $(-\Delta,\Delta)$. As it is in
the corresponding case (i) of Section IV (Fig.~\ref{Fig_9}), small disorder
destroys the band structure of the underlaying periodic system, and with $%
\Delta/\delta$ increasing $\left\langle T(\beta)\right\rangle$ takes, in the
vicinity of the normal incidence (small $\beta)$, a parabolic form, which
remains unchanged when increasing disorder. When $\varepsilon =0$ and $u_j$
is a periodic set of numbers with alternating signs (an array of random p-n
junctions corresponding to the case (iii) of Section IV), the shape of $%
\left\langle T(\beta)\right\rangle$ is also parabolic with the half-width
similar to that shown in Fig.~\ref{Fig_13}.

\section{Conclusions}

We have studied the transport and localization of charge carriers in
graphene superlattices produced by applying periodic and disordered
potentials that depend on one coordinate. Simultaneously, the optical
properties of analogue dielectric structures composed of traditional
(right-handed, RH) dielectric and left-handed (LH) metamaterial layers was
considered and compared with the charge transport in graphene. It was shown
that in the Kronig-Penny-type periodic structures, a sort of total internal
reflection can occur. In the case of a non-symmetric periodic array of
alternating p-n and n-p junctions, along with the conduction bands and band
gaps in the angular domain, there are also a discrete set of directions, in
which the structures are resonantly transparent. In symmetric
($u_{1}=-u_{2}$%
, $\varepsilon =0$, and $\delta _{1}=\delta _{2}$) systems, the conduction
zones disappear, and the angular spectrum of the transmission coefficient
represents a discrete set of resonances, similar to the resonances in the
symmetric ($n_{1}=-n_{2}$, $\delta _{1}=\delta _{2}$, and $Z_{1}\neq Z_{2}$)
periodic alternating RH-LH dielectric structures. These features make the
direction of the charge flux easily tunable by changing the applied voltage.
The numerical experiments have shown that relatively weak disorder can
drastically change the transmission properties of the underlaying periodic
configurations. In the direction orthogonal to the layers created by a 1D
random potential, the eigenstates are extended for all energies and a
minimal conductivity remains non-zero, no matter how strong the disorder is.
Away from the normal incidence, 1D random graphene systems manifest all
features of disorder-induced strong localization. There exist a discrete
random set of angles (or a discrete random set of energies for each given
angle) for which the corresponding wave functions are exponentially
localized. Depending on the type of the unperturbed system, the disorder
could either suppress or enhance the transmission. The transmission of a
graphene system built of alternating p-n and n-p junctions has anomalously
narrow angular spectrum; is extremely sensitive to fluctuations of the
applied potential; and, in some range of directions, it is practically
independent of the amplitude of fluctuations of the potential. Our numerical
results fit well the analytically calculated short wavelength asymptotics of
the mean values of the corresponding transfer matrices. The main features of
the charge transport in graphene subject to a disordered potential have been
compared with those of the propagation of light in inhomogeneous dielectric
media. This comparison has enabled better understanding of both physical
processes.

\section*{Acknowledgments}

FN acknowledges partial support from the National Security Agency (NSA),
Laboratory for Physical Sciences (LPS), Army Research Office (ARO), the
National Science Foundation (NSF) grant No. EIA-0130383, JST, and CREST. FN
and SS acknowledge partial support from JSPS-RFBR 06-02-91200, and
Core-to-Core (CTC) program supported by JSPS. SS acknowledges support from
the Ministry of Science, Culture and Sport of Japan via the Grant-in Aid for
Young Scientists No 18740224, the UK EPSRC via No. EP/D072581/1,
EP/F005482/1, and ESF network-programme ``Arrays of Quantum Dots and
Josephson Junctions''.


\begin{thebibliography}{99}
\bibitem{Katsnelson2} {M.I. Katsnelson and K.S Novoselov, Solid State
Commun. \textbf{143}, 3 (2007).}

\bibitem{Geim} {A.K. Geim and K.S. Novoselov, Nature Mat. \textbf{6}, 183
(2007).}

\bibitem{periodic-1} C-H. Parc, L. Yang, Y-W. Son, M. Cohen, and S. Louie,
Nature Phys. \textbf{4}, 213 (2008).

\bibitem{dis-exp-1} K.S. Novoselov, A.K. Geim, S.V. Morozov, D. Jiang, M.I.
Katsnelson, I.V. Grigorieva, S.V. Dubonos, and A.A. Firsov, Nature (London)
\textbf{438}, 197 (2005).

\bibitem{Castro} {A.H. Castro Neto, F. Guinea, N.M.R. Peres, K.S. Novoselov,
and A.K. Geim, Rev. Mod. Phys. (in press), (2008);
arXiv:cond-mat/0709.1163.}

\bibitem{Katsnelson1} {M.I. Katsnelson, K.S. Novoselov, and A.K. Geim,
Nature Phys. \textbf{2}, 620 (2006).}

\bibitem{Klein} {O. Klein, Z.Phys. \textbf{53}, 157 (1929).}

\bibitem{Cheianov} {V.V. Cheianov, V. Fal'ko, and B.L. Altshuler, Science
\textbf{315}, 1252 (2007).}

\bibitem{Veselago} {V.G. Veselago , Sov. Phys. Uspekhi. \textbf{10}, 509
(1968) [Usp. Fiz. Nauk} \textbf{92}, 517 (1967)].

\bibitem{Pendry} {J.B. Pendry, Phys. Rev. Lett. \textbf{85}, 3966 (2000).}

\bibitem{disorder-1} N.M.R. Peres, F. Guinea, and A.H. Castro Neto, Phys.
Rev. B \textbf{73}, 125411 (2006).

\bibitem{disorder-2} J. Tworzyd\l o, B. Trauzettel, M. Titov, A. Rycerz, and
C.W.J. Beenakker, Phys. Rev. Lett. \textbf{96}, 246802 (2006).

\bibitem{Titov-1} M. Titov, Europhys. Lett. \textbf{79}, 17004 (2007).

\bibitem{San-Jose} P. San-Jose, E. Prada, and D.S. Golubev, Phys. Rev. B
\textbf{76}, 195445 (2007).

\bibitem{Rozhkov} A.V. Rozhkov, S. Savel'ev, and F. Nori,
arXiv:cond-mat/0808.1636.

\bibitem{disorder-3} F. Guinea, M.I. Katsnelson, and M.A.H. Vozmediano,
Phys. Rev. B \textbf{77}, 075422 (2008).

\bibitem{disorder-4} E. Rossi, S. Adam, and S.D. Sarma,
arXiv:cond-mat/0809.1425.

\bibitem{dis-exp-2} Y. Zhang, Y.-W. Tan, H.L. Stormer, and P. Kim, Nature
(London) 438, 201 (2005).

\bibitem{dis-exp-3} X-Z. Yan, C.S. Ting, Phys. Rev. Lett. \textbf{101},
126801 (2008).

\bibitem{remark-1} For comparison of different definitions of localization
and relevant terminology see \cite{math-loc}.

\bibitem{math-loc} C. de Oliveira, R. Prado, J. of Math. Phys. \textbf{46},
072105 (2005).

\bibitem{Slonczewski} {J.C. Slonczewski and P.R. Weiss, Phys. Rev. \textbf{%
109}, 272 (1958).}

\bibitem{Semenoff} {G.W. Semenoff, Phys. Rev. Lett. \textbf{53}, 2449
(1984).%
}

\bibitem{matrices} V. Freilikher, B. Liansky, I. Yurkevich, A. Maradudin,
and A. McGurn, Phys. Rev. E \textbf{51}, 6301 (1995).

\bibitem{Dragoman} D. Dragoman and M. Dragoman, \textit{Quantum-classical
analogies}, Springer, Berlin, (2004).

\bibitem{analogy} P. Darancet, V. Olevano, and D. Mayou,
arXiv:cond-mat/080.3553.

\bibitem{Nori} S. Tamura and F. Nori, Phys. Rev. B \textbf{41}, 7941 (1990).

\bibitem{Nori_1} N. Nishiguchi, S. Tamura, and F. Nori, Phys. Rev. B
\textbf{%
48}, 2515 (1993).

\bibitem{Nori_2} N. Nishiguchi, S. Tamura, and F. Nori, Phys. Rev. B
\textbf{%
48}, 14426 (1993).

\bibitem{born wolf} M. Born and E. Wolf, \textit{Principles of Optics},
Cambridge University Press, Cambridge, UK, (1999).

\bibitem{periodic-2} M. Barbier, F. Peeters, P. Vasilopoulos, and M.
Pereira, Jr, Phys. Rev. B \textbf{77}, 115446 (2008).

\bibitem{periodic-3} C-H Park, L. Yang, Y-W Son, M. Cohen, and S. Louie,
Phys. Rev. Lett. \textbf{101}, 126804 (2008).

\bibitem{periodic-4} T.G. Pedersen, C. Flindt, J. Pedersen, N.A. Mortensen,
A.-P. Jauho, and K. Pedersen, Phys. Rev. Lett. \textbf{100}, 136804 (2008).

\bibitem{periodic-5} M. Diem, T. Koschny, and C. Soukoulis,
arXiv:optics/0807.3351.

\bibitem{Wu1} L. Wu, S. He, and L. Chen, Opt. Express \textbf{11}, 1283
(1983).

\bibitem{Wu2} {L. Wu, S. He, and L. Shen, Phys. Rev. B \textbf{67}, 235103
(2003).}

\bibitem{Dragila} {R. Dragila, B. Luther-Davies, and S. Vukovic, Phys. Rev.
Lett. \textbf{55}, 1117 (1985).}

\bibitem{Pereira} {J.M. Pereira, Jr., V. Mlinar, and F.M. Peeters, Phys.
Rev. B \textbf{74}, 045424 (2006).}

\bibitem{RIKEN} V.A. Yampol'skii, S. Savel'ev, and F. Nori, New J. Phys.
\textbf{10}, 053024 (2008).

\bibitem{Koshino} K. Nomura, M. Koshino, and S. Ryu, Phys. Rev. Lett.
\textbf{99}, 146806 (2007).

\bibitem{Fr-enhancement} V. Freilikher, M. Pustilnik, and I. Yurkevich,
Phys. Rev. B \textbf{53}, 7413 (1996).

\bibitem{Asatryan} A. A. Asatryan, L.C. Botten, M.A. Byrne, V.D. Freilikher,
S.A. Gredeskul, I.V. Shadrivov, R.C. McPhedran, and Y.S. Kivshar, Phys. Rev.
Lett. \textbf{99}, 193902 (2007).

\end{thebibliography}
\end{document}